\documentclass[a4paper,12pt]{article}

\usepackage[T1]{fontenc}
\usepackage[utf8]{inputenc}
\usepackage[english]{babel}
\usepackage{lmodern}

\usepackage{authblk}
\usepackage{amsmath}
\usepackage{amssymb}
\usepackage{cite}
\usepackage{hyperref}
\usepackage{xcolor}
\usepackage{fancyhdr}
\usepackage{tikz}

\renewcommand{\headheight}{42pt}

\title{Factorization of denominators in integration-by-parts reductions}
\author[,a]{Johann Usovitsch%
  \footnote{E-mail: \href{mailto:jusovits@uni-mainz.de}{jusovits@uni-mainz.de}}}
\affil[a]{PRISMA Cluster of Excellence, Institut für Physik\\
Johannes Gutenberg-Universität Mainz\\
D - 55099 Mainz, Germany}
\date{}

\newcommand*{\kira}{\texttt{Kira}}

\newcommand*{\CO}{\mathcal{O}}

\begin{document}

\fancypagestyle{firstpage}{\rhead{
  MITP/20-008\\
}}

\maketitle
\thispagestyle{firstpage}

\begin{abstract}
  We present a \texttt{Mathematica} package which finds a basis of master integrals for the Feynman integral reduction. In this basis the dependence on the dimensional regularization in the denominators factorizes in kinematic independent polynomials, see also \cite{Smirnov:2020quc}.
\end{abstract}

\clearpage
\renewcommand{\headheight}{0pt}

\section{Introduction}

The integration-by-parts (IBP) \cite{Chetyrkin:1981qh} and Lorentz invariance \cite{Gehrmann:1999as} identities are widely used in the calculations of multi-loop scattering amplitudes, especially for physics at the Large Hadron Collider. 
Programs for IBP reductions \cite{Anastasiou:2004vj,Studerus:2009ye,vonManteuffel:2012np,Smirnov:2014hma,Maierhoefer:2017hyi,Smirnov:2019qkx}, which are based on the Laporta algorithm \cite{Laporta:2001dd}, are often the main bottleneck in these calculations.
To keep up with the increasing demands in the applications of Feynman integral reductions, many ideas appeared in recent years including the applications of syzygy equations \cite{Kosower:2018obg,Gluza:2010ws,vonManteuffel:2020vjv,Schabinger:2011dz,Ita:2015tya,Boehm:2017wjc}, algebraic geometry \cite{Larsen:2015ped,Boehm:2018fpv,Bendle:2019csk}, intersection numbers \cite{Mastrolia:2018uzb,Frellesvig:2019kgj,Frellesvig:2019uqt,Weinzierl:2020xyy} and altogether with finite field techniques \cite{vonManteuffel:2014ixa,Peraro:2016wsq,Klappert:2019emp,Peraro:2019svx} or introduction of new integral representation \cite{Wang:2019mnn,Guan:2019bcx,Liu:2018dmc,Liu:2017jxz}. The main goal of these methods and tools is to reduce the computational complexity of the final results appearing in the IBP reductions, which involve problems like scalability of programs with the parallelization of the code, reduction of the total main memory demand and the amount of used disk storage space.

In this paper we describe a method, see also \cite{Smirnov:2020quc}, which reduces the complexity of a reduction problem by choosing a proper master integral basis, such that the dimensional regularization parameter $d$ factorizes in the denominators at the end of the reduction process. In the past a different group made use of this property in \cite{Melnikov:2016qoc} to amplify the reductions. We have implemented this method in the \texttt{Mathematica} package \texttt{findFactorizedBasis.m} and tested it for two non trivial examples: two-loop double pentagon and three-loop non-planar form factor integral.
The performance of the tool \texttt{findFactorizedBasis.m} depends on the underlying reduction program. We have chosen to aid the support with \kira{}.

\section{Algorithm description}

In this section we describe the algorithm, which is implemented in the tool \texttt{findFactorizedBasis.m} to find a new basis of master integrals, where all $d$-dependence of the denominators of the coefficients in the final reduction factorize. We begin with the definition of the Feynman integral in the loop momenta representation:
\begin{align}
 I(\alpha_{1},\dots,\alpha_{N})=\int \prod\limits_{i=1}^{L}\frac{\mathrm{d}^{d}l_{i}}{i\pi^{d/2}}\prod\limits_{j=1}^{N}\frac{1}{D_{j}^{\alpha_{j}}},
 \label{eq:loop-momenta-integral}
\end{align}
where $D_{j}=q_{j}^{2}-m_{j}^{2}$ are the usual inverse propagators, suppressing the Feynman prescription. The momenta $q_{j}$ are linear combination of loop momenta $l_{i}$ and the external momenta $p_{k}$ (or $k_{k}$ both notations are used), $m_{j}$ are the propagator masses and the $\alpha_{j}$ are the propagator powers. We define the sector number of an integral as $S=\sum\limits_{i=1}^{N}2^{i}\,\theta(\alpha_{i}-1/2)$. The top level sector number is the biggest sector number an integral can have when all inverse propagator powers are positive. If some of the inverse propagator powers are zero or negative, the sector numbers are smaller and we call them sub sectors. 
We define the total number of scalar products in the numerator of the integral depending on the loop momenta $s=\sum_{i=1}^{N}\alpha_{i}\;\theta(-\alpha_{i}+1/2)$ and the sum of all positive propagator powers $r=\sum_{i=1}^{N}\alpha_{i}\;\theta(\alpha_{i}-1/2)$. Another important number is the number of dots which is: $\sum_{i=1}^{N}\alpha_{i}\;\theta(\alpha_{i}-3/2)$. 

An integral in Eq.~\eqref{eq:loop-momenta-integral} can be expressed in terms of so called minimal basis of master integrals: $I_{j}=\sum_{i}^{N_{m}}C_{ij}(d)M_{i}$, where $N_{m}$ is the total number of master integrals, the coefficients $C_{ij}=\frac{N_{ij}}{D_{ij}}$ are rational functions in $d$, and $M_{i}$ are the master integrals. The functions $N_{ij}$ and $D_{ij}$ are polynomials in $d$ and the kinematic invariants and masses.
Factorization gives $N_{ij}=\prod_{k=1}^{N_{d}}n_{ijk}^{\mu_{ijk}}$ and $D_{ij}=\prod_{k=1}^{N_{d}}d_{ijk}^{\nu_{ijk}}$, where $n_{ijk}$ and $d_{ijk}$ are again polynomials raised to some integer power $\mu_{ijk}$ and $\nu_{ijk}$.
With these definitions at hand we are now ready to go through the aforementioned algorithm:
\begin{itemize}
 \item[1] We perform two reductions $a=1,2$ for all integrals $\{I^{a}_{1},\dots,I^{a}_{N_{r}}\}$ in the top level sector and its all sub sectors including integrals with up to two dots and two scalar products \footnote{All master integrals must appear in the reduction and if all integrals with one dot are master integrals, then integrals with two dots must be included into the reduction.}. 
 We use for each reduction $a$ different numeric values for all scales $\{\dots,s_{i},\dots,m_{j},\dots\}$, where $s_{i}$ are kinematic invariants and $m_{j}$ are the masses in the propagators. The functions $n_{ijk}$ and $d_{ijk}$ are polynomials in $d$ only.
 \item[2] We take the most complicated non zero integral in the Laporta ordering $I_{N_{r}}$ expressed in terms of the master integrals and collect all different building blocks $d^{a}_{ijk}$. We compare all building blocks from different reductions $a=1,2$. The polynomials $d^{a}_{ijk}$ must be functions of the kinematic invariants and masses if $d^{1}_{ijk}\neq{}d^{2}_{ijk}$ modulo sign. We collect these building blocks into the set $R$.
 \item[3] In the first iteration the sector $S$ is the top level sector. 
 We collect all terms in $I_{N_{r}}$ of which master integrals belong to the sector $S$ in the set $P_{S}$.
 We replace one master integral $M_{p}$ in the set $P_{S}$ by a new master integral $M_{q}$ from the set $Q_{S}$. The set $Q_{S}$ contains all integrals of sector $S$ with up to two dots. We check whether the new set of terms in $P'_{S}$ contains a denominator $d'_{N_{r}jk}$ which belongs to the set $R$. 
 If yes, then we take $x_{qp}=\min(\max (\deg (\{n'_{N_{r}1k}\})),\dots,\max(\deg (\{n'_{N_{r}N_{p}k}\})))$ and repeat it for all possible combinations of $q$ and $p$. We collect all possible replacements of master integrals which have the same smallest $x_{pq}$. 
 If not, we go to step 1 and proceed with the sub sectors.
 Finally if one of the $x_{pq}$ replacements involves only kinematic independent $\{n'_{N_{r}jk}\}$ polynomials, then we take this replacement, otherwise take any other replacement with the smallest $x_{pq}$. 
 \item[4] We repeat step 1 to 3 until the set $R$ is empty.
\end{itemize}

\section{Obtainining and setting up\\ \texttt{findFactorizedBasis.m}}

To obtain the latest release version of \texttt{Mathematica} package\\ \texttt{findFactorizedBasis.m}, clone the repository with
\begin{verbatim}
  git clone https://gitlab.com/jusovitsch/findfactorizedbasis.git
\end{verbatim}
checking out the master branch. This version is tested only under Linux.

The tool \texttt{findFactorizedBasis.m} is only compatible with the most recent version of \kira{} from \url{https://gitlab.com/kira-pyred/kira}. \kira{} is a \texttt{C++} program. To run this package one should prepare a working directory containing a \kira{} job file and the config files \texttt{integralfamilies.yaml} and \texttt{kinematics.yaml}, for example:
\begin{verbatim}
#jobs.yaml
jobs:
 - reduce_sectors:
    reduce:
     - {sectors: [255], r: 9, s: 2}
    select_integrals:
      select_mandatory_recursively:
       - {sectors: [255], r: 9, s: 2, d: 1}
    preferred_masters: preferred
    select_masters_reduction:
      - trimBasis
    run_initiate: true
    run_triangular: sectorwise
    run_back_substitution: true
    conditional: true
 - kira2math:
    target:
     - {sectors: [255], r: 9, s: 2, d: 1}
\end{verbatim}

One should adjust the following lines such that all master integrals appear in the reduction:
\begin{verbatim}
{sectors: [255], r: 9, s: 2}
\end{verbatim}
and
\begin{verbatim}
{sectors: [255], r: 9, s: 2, d: 1}
\end{verbatim}
We make use in the job file of the following option:
\begin{verbatim}
select_masters_reduction:
  - trimBasis
\end{verbatim}
With this option \kira{} reads a file named \texttt{trimBasis} containing for example some integral appearing in the reduction:
\begin{verbatim}
#trimBasis
 doublePentagon[1,1,1,1,1,1,1,1,1,0,0]
\end{verbatim}
This option sets all sectors to zero which do not have any dependence to the sector, which the integral \texttt{doublePentagon[1,1,1,1,1,1,1,1,1,0,0]} belongs to. This option is based on the same ideas as it was presented in \kira{} release notes \cite{Maierhofer:2018gpa} and in \cite{Chawdhry:2018awn}. To use \texttt{findFactorizedBasis.m} we do not need to create a file \texttt{trimBasis}, this tool creates it automatically.

The \kira{} \texttt{config} files are generated as documented in the original \kira{} paper \cite{Maierhoefer:2017hyi}, for example:
\begin{verbatim}
integralfamilies:
  - name: "doublePentagon"
    loop_momenta: [l1, l2]
    top_level_sectors: [255]
    propagators:
      - [ "l1", 0 ]             #1
      - [ "l1-k1", 0 ]          #2
      - [ "l1-k1-k2", 0 ]       #3
      - [ "l2", 0 ]             #4
      - [ "l2-k1-k2-k3", 0 ]    #5
      - [ "l2-k1-k2-k3-k4", 0 ] #6
      - [ "l1-l2", 0 ]          #7
      - [ "l1-l2+k3", 0 ]       #8
      - [ "l1-k1-k2-k3-k4", 0 ] #9     
      - [ "l2-k1", 0 ]          #10
      - [ "l2-k1-k2", 0 ]       #11
\end{verbatim}
and
\begin{verbatim}
kinematics :
  incoming_momenta: []
  outgoing_momenta: [k1, k2, k3, k4, k5]
  momentum_conservation: [k5,-k1-k2-k3-k4]
  kinematic_invariants:
    - [s12,  2]
    - [s23,  2]
    - [s34,  2]
    - [s15,  2]
    - [s45,  2]
  scalarproduct_rules:
    - [[k1,k1],  0]
    - [[k2,k2],  0]
    - [[k3,k3],  0]
    - [[k4,k4],  0]
    - [[k5,k5],  0]
    - [[k1+k2,k1+k2],  "s12"]
    - [[k1+k3,k1+k3],  "s45-s12-s23"]
    - [[k1+k4,k1+k4],  "s23-s15-s45"]
    - [[k2+k3,k2+k3],  "s23"]
    - [[k2+k4,k2+k4],  "s15-s23-s34"]
    - [[k3+k4,k3+k4],  "s34"]
  symbol_to_replace_by_one: s12
\end{verbatim}
The loop momenta rooting and the kinematics definition has no impact to this paper and may be chosen arbitrary.

\section{Run \texttt{findFactorizedBasis.m}}

The tool \texttt{findFactorizedBasis.m} completely automates the above algorithm up to the organization of the working directory for \kira{}.

To run the tool we need to create e.g. a \texttt{wolframscript} or we may run it from the \texttt{Mathematica} notebook itself:
\begin{verbatim}
#!/usr/bin/env wolframscript
<< "../../findFactorizedBasis.m"
getNewBasis[doublePentagon, {s23, s45, s15, s34}, \
"../../bin/kira -i2 jobs.yaml -p32", d, 10]
\end{verbatim}
Here we assume that the package \texttt{findFactorizedBasis.m} is two directories above. The last line launches the algorithm described before. The arguments of the function \texttt{getNewBasis} are the following from left to right: \texttt{doublePentagon} is the name of the topology also used in the config file \texttt{integralfamilies.yaml}, \texttt{\{s23, s45, s15, s34\}} is a set of scales which appear in the reduction and are as defined in the config file \texttt{kinematics.yaml}, \texttt{"path/to/kira/executable -i2 jobs.yaml -p32"} is the usual command line option to run \kira{}. Here we instruct \kira{} to run 32 Fermat jobs with \texttt{-p32} and to use the integral ordering only dots with \texttt{-i2}. Here \texttt{jobs.yaml} is the job file. The argument \texttt{d} is the name for the variable which should be factored out in the denominators. The last numeric argument \texttt{10} is optional and can be any positive integer, it changes the kinematic sample points for the scales appearing in the reduction problem.

The new preferred basis of master integrals is written to the file \texttt{preferred} in the working directory of \kira{}.

Furthermore \texttt{findFactorizedBasis.m} writes to backup files: \texttt{resultsA.m} \texttt{resultsB.m}, \texttt{problematicSectors} and \texttt{trimBasis}. If the function\\ \texttt{getNewBasis} unexpectedly terminates, the whole process can be resumed by invoking \texttt{getNewBasis} again. 

We remark that providing the option \texttt{-i2}, \texttt{getNewBasis} terminates successfully faster than with the option \texttt{-i1} (which is default in \kira{}).

The instruction is to use the most up to date version of \kira{} due to the introduction of three new options in \kira{}. The option:\\ \texttt{select\_masters\_reduction: [trimBasis]} was introduced above. Two more options are the following command line options in \kira{}:\\
\texttt{----set\_value=s12=1}\\
and\\
\texttt{----set\_sector=127}. The option \texttt{set\_sector} is used by the program\\ \texttt{findFactorizedBasis.m} to replace the top level sector defined in the \kira{} config files by a new sector. The function \texttt{getNewBasis} uses automatically the option \texttt{set\_value} to set variables which were introduced in\\ \texttt{kinematics.yaml} to some specific numeric values.

\section{Examples}

\subsection{Three-loop non-planar}
\begin{figure}[htpb]
  \begin{center}
    \includegraphics[width=0.6\linewidth]{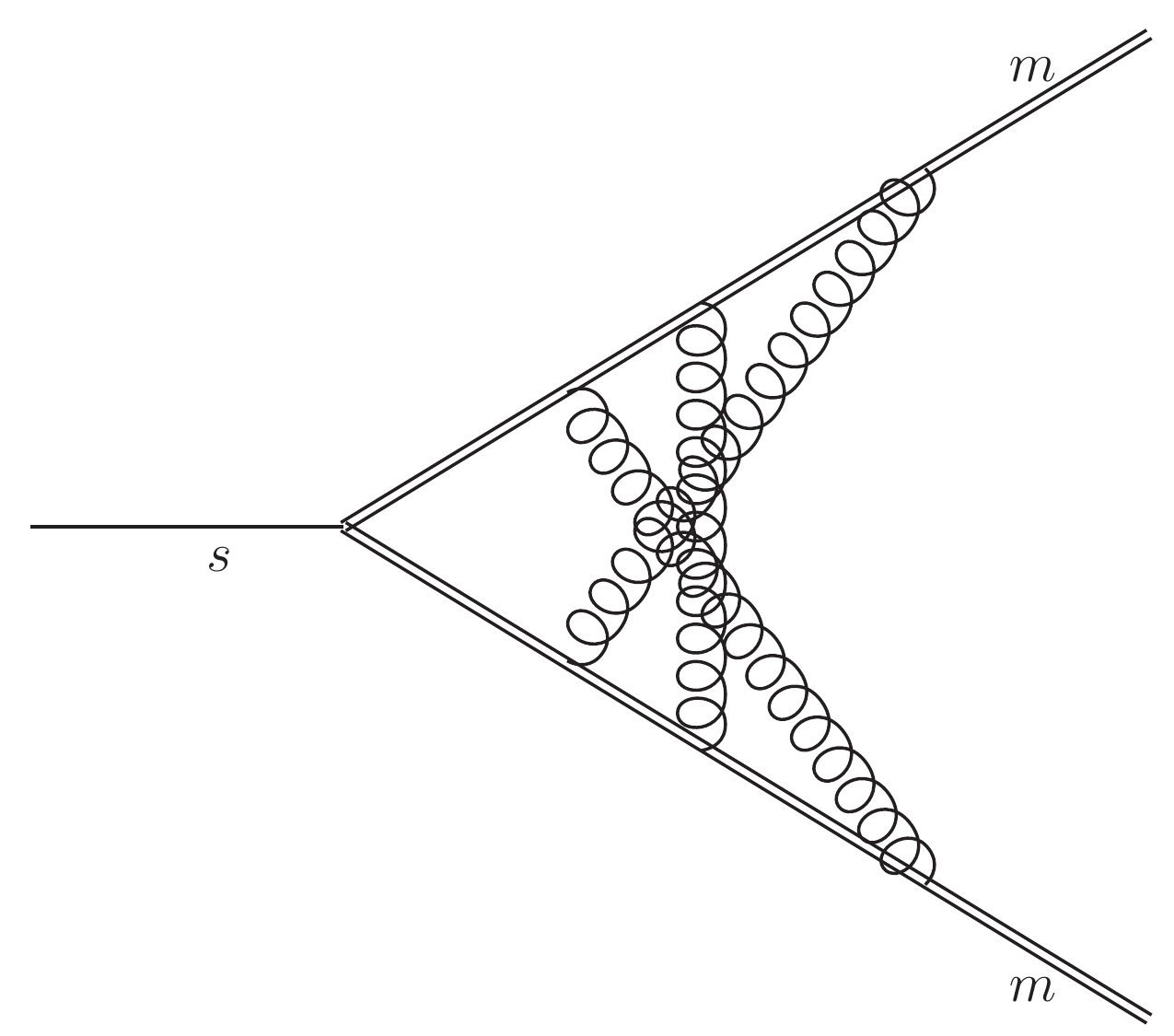}
  \end{center}
   \caption{Three-loop Feynman diagram for the top level sector of an integral in Eq.~\eqref{eq:threeLoop}.}
     \label{fig:threeLoop}
\end{figure}
The three-loop Feynman integral, see Fig.~\ref{fig:threeLoop} is defined as:
\begin{align}
 I^{\texttt{threeLoop}}(\alpha_{1}, \dots , \alpha_{12}) = \int\prod\limits_{i=1}^{3}\frac{\mathrm{d}^{d}l_{i}}{i\pi^{d/2}}\frac{D_{10}^{-\alpha_{10}}D_{11}^{-\alpha_{11}}D_{12}^{-\alpha_{12}}}{D_{1}^{\alpha_{1}}D_{2}^{\alpha_{2}}D_{3}^{\alpha_{3}}D_{4}^{\alpha_{4}}D_{5}^{\alpha_{5}}D_{6}^{\alpha_{6}}D_{7}^{\alpha_{7}}D_{8}^{\alpha_{8}}D_{9}^{\alpha_{9}}},
 \label{eq:threeLoop}
\end{align}
with the inverse propagators:
\begin{align}
&D_{1}=l_{1}^{2}-m, D_{2}=l_{2}^{2}-m,\nonumber\\
&D_{3}=l_{3}^{2}-m, D_{4}=(l_{1}-p_{1})^{2} - m,\nonumber\\
&D_{5}=(l_{1}-l_{3}+p_{2})^2 - m, D_{6}=(l_{1}+l_{2}+p_{2})^2 - m,\nonumber\\
&D_{7}=(l_{1}+l_{2})^{2}, D_{8}=(l_{3}-p_{2}-p_{1})^{2},\nonumber\\
&D_{9}=(l_{2}+l_{3})^{2}, 
\end{align}
and the auxiliary propagators are:
\begin{align}
D_{10}=l_{1}l_{3}, D_{11}=l_{2}p_{1}, D_{12}=l_{2}p_{2}.  
\end{align}

The kinematics are: $p_{1}^{2}=s=1$ and $(p_{1}+p_{2})^{2}=p_{2}^{2}=p_{3}^{2}=m$. The variable $m$ is the squared mass of the inverse propagator. 
Any integral with $\alpha_{i}\geq0$ can be written in terms of 159 master integrals.
The program \texttt{findFactorizedBasis.m} with \kira{} using the option \texttt{integral\_ordering: 2} finds a preferred list of master integrals, which can be found in the file \texttt{preferred} in the example directory \texttt{example/threeLoop}. The $d$-dependence completely factorizes for all denominators for any integral reduced to this basis. 

Now we could perform the reduction to this basis for any Feynman integral by setting first $m$ to some numeric value, collect all denominators which are polynomials in $d$. After that we could repeat the reduction by setting $d$ to some numeric value and collect the denominators as a function of the scale $m$. Put together the denominator polynomials depending on $d$ and $m$ from both reductions. That way we are able to get the analytic structure for any denominator for any integral. Note that all reductions are performed with one variable less compared to the full reduction, where full means no variables are set to numeric values. After this we can proceed with the full reduction, this time canceling the denominators beforehand. 

Another main feature is that the denominators in this basis factorize in few polynomials of $\CO(10-100)$ of low degree raised to integer powers. This safes the disk space by a factor of 2, compared to a reduction without the factorization. Furthermore, if we perform the full reduction with finite field methods with the canceling of the denominators we need just polynomial reconstruction, which needs less samplings and is in general simpler.

\subsection{Double pentagon}
\begin{figure}[htpb]
  \begin{center}
    \includegraphics{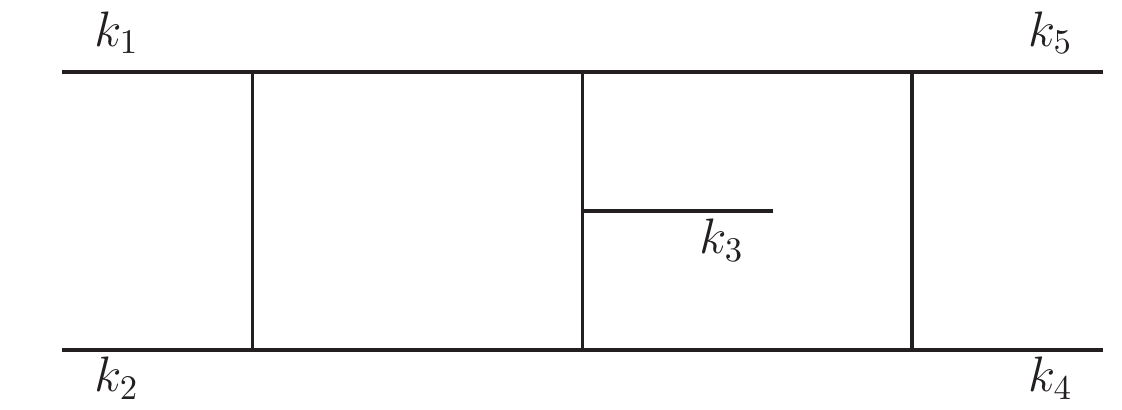}
  \end{center}
   \caption{This is a Feynman diagram for one top level sector integral for the double pentagon.}
   \label{fig:doublePentagon}
\end{figure}
The double pentagon Feynman integral is defined as:
\begin{align}
 I^{\texttt{doublePentagon}}(\alpha_{1},\dots,\alpha_{11})=\int\prod\limits_{i=1}^{2}\frac{\mathrm{d}^{d}l_{i}}{i\pi^{d/2}}\frac{D_{9}^{-\alpha_{9}}D_{10}^{-\alpha_{10}}D_{11}^{-\alpha_{11}}}{D_{1}^{\alpha_{1}}D_{2}^{\alpha_{2}}D_{3}^{\alpha_{3}}D_{4}^{\alpha_{4}}D_{5}^{\alpha_{5}}D_{6}^{\alpha_{6}}D_{7}^{\alpha_{7}}D_{8}^{\alpha_{8}}},
 \label{eq:doublePentagon}
\end{align}
with the inverse propagators:
\begin{align}
&D_{1}=l_{1}^2, D_{2}=(l_{1}-k_{1})^2,\nonumber\\
&D_{3}=(l_{1}-k_{1}-k_{2})^2, D_{4}=l_{2}^2,\nonumber\\
&D_{5}=(l_{2}-k_{1}-k_{2}-k_{3})^2, D_{6}=(l_{2}-k_{1}-k_{2}-k_{3}-k_{4})^2,\nonumber\\
&D_{7}=(l_{1}-l_{2})^2, D_{8}=(l_{1}-l_{2}+k_{3})^2,
\end{align}
and the auxiliary propagators are: 
\begin{align}
D_{9}=(l_{2}-k_{1}-k_{2})^2, D_{10}=(l_{1}-k_{1}-k_{2}-k_{3}-k_{4})^2, D_{11}=(l_{2}-k_{1})^2. 
\end{align}
The kinematics are chosen as:
\begin{align}
&k_{1}k_{2}=\frac{s_{12}}{2}, k_{1}k_{3}=\frac{s_{45}-s_{12}-s_{23}}{2},\\ &k_{1}k_{4}=\frac{s_{23}-s_{15}-s_{45}}{2}, k_{2}k_{3}=\frac{s_{23}}{2},\\ &k_{2}k_{4}=\frac{s_{15}-s_{23}-s_{34}}{2}, k_{3}k_{4}=\frac{s_{34}}{2},\\ &k_{1}^{2}=k_{2}^{2}=k_{3}^{2}=k_{4}^{2}=k_{5}^{2}=0, s_{12}=1.
\end{align}
The reduction of any integral $\alpha_{i}\geq0$ gives total number of master integrals: 108.
 
The tool \texttt{findFactorizedBasis.m} generates automatically a preferred basis of master integrals which factorizes the $d$-dependence in the denominators, see file \texttt{preferred} in the example directory \texttt{example/doublePentagon}.
 
In further discussion we use an alternative basis, see file \texttt{preferred} in \texttt{example/doublePentagon2}, which is of the same complexity as the previous one.
We determined all denominators for all coefficients in the final reduction for the following 21 integrals for the topology, see Fig.~\ref{fig:doublePentagon}:
\begin{align}
I_{1}&=I^{\texttt{doublePentagon}}(1,1,1,1,1,1,1,1,-5,0,0),\nonumber\\
I_{2}&=I^{\texttt{doublePentagon}}(1,1,1,1,1,1,1,1, 0,-5,0),\nonumber\\
I_{3}&=I^{\texttt{doublePentagon}}(1,1,1,1,1,1,1,1, 0,0,-5),\nonumber\\
I_{4}&=I^{\texttt{doublePentagon}}(1,1,1,1,1,1,1,1,-4,-1,0),\nonumber\\
I_{5}&=I^{\texttt{doublePentagon}}(1,1,1,1,1,1,1,1,-4,0,-1),\nonumber\\
I_{6}&=I^{\texttt{doublePentagon}}(1,1,1,1,1,1,1,1,-1,-4,0),\nonumber\\
I_{7}&=I^{\texttt{doublePentagon}}(1,1,1,1,1,1,1,1, 0,-4,-1),\nonumber\\
I_{8}&=I^{\texttt{doublePentagon}}(1,1,1,1,1,1,1,1,-1,0,-4),\nonumber\\
I_{9}&=I^{\texttt{doublePentagon}}(1,1,1,1,1,1,1,1, 0,-1,-4),\nonumber\\
I_{10}&=I^{\texttt{doublePentagon}}(1,1,1,1,1,1,1,1,-3,-2,0),\nonumber\\
I_{11}&=I^{\texttt{doublePentagon}}(1,1,1,1,1,1,1,1,-3,0,-2),\nonumber\\
I_{12}&=I^{\texttt{doublePentagon}}(1,1,1,1,1,1,1,1,-3,-1,-1),\nonumber\\
I_{13}&=I^{\texttt{doublePentagon}}(1,1,1,1,1,1,1,1,-2,-3,0),\nonumber\\
I_{14}&=I^{\texttt{doublePentagon}}(1,1,1,1,1,1,1,1, 0,-3,-2),\nonumber\\
I_{15}&=I^{\texttt{doublePentagon}}(1,1,1,1,1,1,1,1,-1,-3,-1),\nonumber\\
I_{16}&=I^{\texttt{doublePentagon}}(1,1,1,1,1,1,1,1,-2,0,-3),\nonumber\\
I_{17}&=I^{\texttt{doublePentagon}}(1,1,1,1,1,1,1,1, 0,-2,-3),\nonumber\\
I_{18}&=I^{\texttt{doublePentagon}}(1,1,1,1,1,1,1,1,-1,-1,-3),\nonumber\\
I_{19}&=I^{\texttt{doublePentagon}}(1,1,1,1,1,1,1,1,-2,-2,-1),\nonumber\\
I_{20}&=I^{\texttt{doublePentagon}}(1,1,1,1,1,1,1,1,-2,-1,-2),\nonumber\\
I_{21}&=I^{\texttt{doublePentagon}}(1,1,1,1,1,1,1,1,-1,-2,-2).
\end{align}
We parametrize our result, see Tab.~\ref{tab:doublePentagon} in the following way:
\begin{align}
 I_{j}&=\sum_{k}C_{jk}M_{k},\quad C_{jk}=\frac{n_{jk}}{d_{jk}},\nonumber\\
 d_{jk}&=\prod_{l}d_{jkl}(d)^{\nu_{jkl}}\prod_{m}d_{jkm}(\{s_{12},s_{23},s_{34},s_{15},s_{45}\})^{\nu_{jkm}},
 \label{eq:buildingblocks}
\end{align}
where we set $s_{12}=1$.

\begin{table}[htpb]
\renewcommand{\arraystretch}{1.2}
\caption{Denominator building blocks, see Eq.~\eqref{eq:buildingblocks} are sufficient to express all denominators in the reduction of all 21 integrals with 5 scalar products.}
\begin{center}
\begin{tabular}{ll}
\hline
 $d_{jkl}$	& $d_{jkm}$\\\hline
$d - 8$		&     $(s_{15} - s_{23} + s_{45})$\\
$d - 6$		& $(-1 - s_{15} + s_{34})$\\
$d - 5$		&     $(-s_{15} + s_{23} - s_{45})$\\
$d - 4$		& $(-1 - s_{15} + s_{34})$\\
$d - 3$		& $(-1 + s_{34})$\\
$d - 2$		&     $1 + s_{23} - s_{45}$\\
$d - 1$		&     $(-s_{15} + s_{23} + s_{34})$\\
$2 d - 11$	&$(-1 + s_{34} + s_{45})$\\
$2 d - 9$	&$(s_{34} + s_{45})$\\
$2 d - 7$	&     $(s_{15}^2 - 2 s_{15} s_{23} + s_{23}^2 + 2 s_{15} s_{23} s_{34} - 2 s_{23}^2 s_{34} + s_{23}^2 s_{34}^2$ \\
		&     $- 2 s_{15}^2 s_{45} + 2 s_{15} s_{23} s_{45} + 2 s_{15} s_{34} s_{45} + 2 s_{23} s_{34} s_{45}$\\
		&     $+ 2 s_{15} s_{23} s_{34} s_{45} - 2 s_{23} s_{34}^2 s_{45} + s_{15}^2 s_{45}^2 - 2 s_{15} s_{34} s_{45}^2 + s_{34}^2 s_{45}^2)$\\
$3 d - 10$	&     $(s_{15} s_{34} - s_{23} s_{34} + s_{23} s_{34}^2 - s_{15} s_{45} + s_{23} s_{45} + 2 s_{34} s_{45}$\\
		&     $+ s_{15} s_{34} s_{45} + s_{23} s_{34} s_{45} - s_{34}^2 s_{45} + s_{15} s_{45}^2 - s_{34} s_{45}^2)$\\
$3 d - 8$	& $(1 + s_{23})$\\
$3 d - 14$	& $(-s_{15} + s_{34})$\\
		& $(s_{23})$, $(s_{15})$, $(s_{45})$, $(s_{34})$\\
		&  $(1 + s_{23} - s_{34} - 2 s_{23} s_{34} - 2 s_{45} - s_{23} s_{45} + s_{34} s_{45} + s_{45}^2)$\\
		&   $(-1 + s_{34} + s_{45} + s_{34} s_{45})$\\
		&$(s_{15} - s_{23})$\\
		&$(s_{23} + s_{34})$\\
		&   $(-s_{15} + s_{23} - s_{23} s_{34} - s_{45} + s_{15} s_{45} - 2 s_{23} s_{45} + s_{45}^2)$\\
		&  $(-s_{23} - s_{45} - s_{23} s_{45} + s_{45}^2)$\\
		&   $(-1 + s_{15} - s_{23} + s_{45})$\\
		&$(s_{23} - s_{45})$\\
		&$(1 + s_{15} - s_{34} - s_{45})$\\
		&  $(-1 + s_{45})$\\
		&$(s_{15} - s_{23} + s_{23} s_{34} - s_{15} s_{45} + s_{34} s_{45})$\\
\hline
\end{tabular}
\label{tab:doublePentagon}
\end{center}
\end{table}
\clearpage

\section{Summary}
We have given an introduction to a new tool \texttt{findFactorizedBasis.m} which automatically generates a basis of master integrals, such that all $d$-dependence in the denominators factorize. We have mentioned several advantages using the basis generated by the tool \texttt{findFactorizedBasis.m}. For example: the result tables containing the final IBP reduction are halved in size, since denominators factorize into few polynomials of low degree raised to integer powers.
We demonstrated that the analytic structures of the denominators are straightforward to determine with the methods of algebraic reconstruction. For algebraic reconstruction we used just few numeric samples of the full reduction with all variables set to numeric values but one. We encourage the use of \texttt{findFactorizedBasis.m} together with IBP reductions based on the finite field methods. Because once we know the analytic structure of the denominators, we can perform the full reduction just for the numerators. That way, the computation time is at least halved and one needs algorithms just for the reconstruction of the polynomials.
Our figures were generated using \texttt{Jaxodraw}\cite{Binosi:2003yf}, based on \texttt{AxoDraw} \cite{Vermaseren:1994je}.

\section*{Acknowledgments}

We thank Alexander Smirnov and Vladimir Smirnov for their share of knowledge and ideas and for fair competition. We thank Fabian Lange and Jonas Klappert for their independent verification of the denominators for the 21 integrals belonging to the double pentagon.
We thank Matthias Steinhauser for the suggestion to study the three-loop example, which is presented in this paper.
We thank Zoltán Szőr for proof reading this manuscript.

Parts of this research were conducted using the supercomputer Mogon and/or advisory services offered by Johannes Gutenberg University Mainz (hpc.uni-mainz.de), which is a member of the AHRP (Alliance for High Performance Computing in Rhineland Palatinate,  www.ahrp.info) and the Gauss Alliance e.V.

The authors gratefully acknowledge the computing time granted on the supercomputer Mogon at Johannes Gutenberg University Mainz (hpc.uni-mainz.de).


\clearpage

\bibliographystyle{JHEP}

\providecommand{\href}[2]{#2}\begingroup\raggedright\endgroup

\end{document}